.

PROJET STS

# L'Ordinateur Quantique

Denis Crottet

EPFL, Physique, 8ème sem.

Ecublens, le 29 mars 2000

# Table des matières

# Chapitre 1

# Introduction

A l'heure actuelle, l'informatique a pris une place considérable, si ce n'est essentielle, dans le monde de l'industrie et des sciences. Dans le but d'être toujours plus rapides et toujours plus performants, les différents utilisateurs demandent un matériel informatique adapté à leurs besoins. Ainsi, nous assistons à une incessante amélioration de la capacité des processeurs, des mémoires et des périphériques composant un ordinateur. Le nombre de transistors qu'il est possible de placer sur un chip croît exponentiellement dans le temps, un constat appelé "Loi de Moore". Dans ce contexte, l'idée de changer le système informatique de base actuel au lieu de l'améliorer constamment est bien entendu prise en compte. L'une de ces propositions est l'ordinateur quantique.

Historiquement, c'est en 1985 que D. Deutsch démontra théoriquement qu'il était possible de réaliser une machine de Turing à l'aide des concepts de mécanique quantique [1]. Le premier pas de la recherche pour un ordinateur quantique était fait. Cependant, l'explosion de la recherche dans ce domaine eut lieu en 1994, lorsque P. Shor proposa un algorithme pour déterminer la période d'une fonction donnée (étape essentielle pour la factorisation d'un nombre) [2]. Ce dernier est un algorithme polynomial (le nombre d'étape varie comme $N^\alpha$), alors qu'il n'en existe pas d'aussi puissant lors d'une programmation classique. Par la suite d'autres algorithmes avec un nombre d'étapes réduites par rapport aux algorithmes classiques ont été démontrés. Citons l'algorithme de Grover qui permet de sélectionner un élément d'une liste [3] et celui de Deutsch-Jozsa qui permet de déterminer si une fonction donnée $f$ est constante ou non [4]. Il est donc, à première vue, intéressant de chercher à réaliser un ordinateur quantique. L'artricle de revue le plus récent est celui de C.H. Bennett et D.P. DiVincenzo [5]

Dans cette étude, après avoir décrit le concept de l'ordinateur quantique, une analyse critique des premières réalisations expérimentales est présentée. Avant de conclure, un commentaire sur l'importance que l'ordinateur quantique a prise dans la recherche en physique depuis 1994 sera également donné.

# Chapitre 2

# Le concept de l'ordinateur quantique

## 2.1 Rappel de concepts de mécanique quantique

### 2.1.1 Description du système : l'espace de Hilbert

Lorsque nous étudions un système physique à l'aide des lois de la mécanique quantique, nous travaillons dans un espace mathématique appelé espace de Hilbert, car ce dernier possède les propriétés nécessaires (linéarité, existence d'un produit scalaire et complétude) pour satisfaire les principes de base de la théorie. Si nous prenons un système $S$ composé de sous-systèmes $S_1$, ..., $S_n$, l'espace de Hilbert correspondant est construit comme l'espace produit tensoriel des espaces décrivant chaque sous-système : $\mathcal{H} = \mathcal{H}_1 \otimes \ldots \otimes \mathcal{H}_n$. Ainsi, en identifiant un vecteur $|\psi\rangle \in \mathcal{H}$ à un état du système, nous nous rendons compte, de par la construction mathématique, de l'existence d'états de superposition. En effet, grâce à la linéarité de l'espace nous avons : $a|\psi\rangle + b|\phi\rangle \in \mathcal{H}$, $\forall \psi, \phi \in \mathcal{H}$ et $\forall a, b \in \mathbb{C}$. Remarquons que même lorsque $|\psi\rangle = |\psi_1\rangle \otimes \ldots \otimes |\psi_n\rangle$, $|\phi\rangle = |\phi_1\rangle \otimes \ldots \otimes |\phi_n\rangle$ il n'est en général pas possible d'écrire l'état de superposition comme un produit tensoriel d'états des sous-systèmes: il existe ainsi dans $\mathcal{H}$ des états non-factorisables, appelés aussi états intriqués (entangled states). L'état intriqué décrit une situation dans laquelle, bien que l'état du système global $S$ soit défini, l'état de chaque sous-système $S_i$, ne l'est pas. Pour illustrer ce concept, prenons comme exemple un système $S$ composé de deux sous-systèmes $S_1$ et $S_2$ à deux niveaux (notés $+$ et $-$) :

$$\begin{aligned} 1) \quad & |\psi\rangle = |--\rangle + |-+\rangle = |-\rangle \otimes [|-\rangle + |+\rangle] \quad \text{état factorisable} \\ 2) \quad & |\psi\rangle = |--\rangle + |++\rangle \neq |\psi_1\rangle \otimes |\psi_2\rangle \quad \text{état intriqué.} \end{aligned} \quad (2.1)$$

Un système comportant un nombre fini $m$ d'états orthogonaux est décrit par un espace de Hilbert complexe de dimension $m$. Cet espace est donc isomorphe à $\mathbb{C}^m$. Comme nous le discuterons dans la prochaine section, pour des ressemblances avec le langage informatique classique il est habituel, lorsque nous traitons l'ordinateur quantique, de considérer comme système quantique typique un système formé de $n$ systèmes à deux niveaux, décrit donc par

$$\mathcal{H} = \underbrace{\mathbb{C}^2 \otimes \ldots \otimes \mathbb{C}^2}_{n \text{ fois}} \simeq \mathbb{C}^{2^n}. \tag{2.2}$$

Il est clair que tout système décrit par $\mathbb{C}^m$ peut être mis en correspondance avec un sous-espace de $\mathbb{C}^{2^n}$ pour $m \leq 2^n < m+1$. Le fait que tous les systèmes quantiques possédant le même nombre d'états orthogonaux soient décrits par le même espace vectoriel a inspiré l'une des premières réflexions sur l'ordinateur quantique, due à Feynman[1].

### 2.1.2 Evolution : opérateur unitaire

En mécanique quantique, l'évolution du système est régie par un opérateur unitaire, généralement noté $U$ (i.e. $U^{-1} = U^+$). Remarquons qu'un opérateur unitaire conserve le produit scalaire :

$$\langle U\psi|U\phi\rangle = \langle U^+U\psi|\phi\rangle = \langle U^{-1}U\psi|\phi\rangle = \langle \psi|\phi\rangle. \tag{2.3}$$

Une discussion des hypothèses qui mènent à adopter un opérateur d'évolution unitaire se trouve dans [7] par. 8.6, p. 237. L'équation de Schrödinger est une conséquence assez immédiate de l'unitarité de l'évolution.

### 2.1.3 Couplage avec l'environnement ("décohérence")

Pour que la théorie rende compte de l'observation, il est nécessaire d'introduire le couplage du système avec son environnement. Par la préparation, nous plaçons le système dans un état bien défini $|\phi\rangle$. L'état initial de l'environnement est $|e\rangle$. S'il y a couplage entre l'environnement est le système (i.e., si l'opérateur d'évolution n'est pas de la forme $U_s \otimes U_e$), on parle d'évolution "bruitée". Algébriquement, cette situation se décrit de la manière suivante :

$$\begin{array}{c} |\phi\rangle \otimes |e\rangle \\ \downarrow \qquad \text{Evolution unitaire bruitée} \\ \sum_i c_i |\phi_i\rangle \otimes |e_i\rangle. \end{array} \tag{2.4}$$

Vu le nombre énorme de degrés de liberté de l'environnement, il est naturel d'admettre que — si le temps écoulé depuis la préparation est assez long — les $|e_i\rangle$ sont orthogonaux entre eux. Ainsi,

---

[1] Feynman [6] envisage la possibilité de simuler le comportement d'un système quantique en utilisant non pas un ordinateur (classique), mais *un autre système quantique* qu'on pourrait contrôler.

en prenant la moyenne d'une observable du système $\mathcal{A} = A \otimes \mathbb{1}$ nous obtenons :

$$\begin{aligned}\langle\psi|\mathcal{A}|\psi\rangle &= \sum_i |c_i|^2 \langle\psi_i|A|\psi_i\rangle \underbrace{\langle e_i|e_i\rangle}_{=1} + \sum_{i\neq j} c_i^* c_j \langle\psi_i|A|\psi_j\rangle \underbrace{\langle e_i|e_j\rangle}_{=0} \\ &= \sum_i |c_i|^2 \langle\psi_i|A|\psi_i\rangle\end{aligned} \qquad (2.5)$$

De par le couplage avec l'environnement, les termes croisés sont nuls. Ceci n'est évidemment plus vrai si nous considérons une évolution non bruitée (i.e. $U = U_s \otimes U_e$) :

$$\begin{aligned}&|\phi\rangle \otimes |e\rangle \\ &\quad\downarrow \qquad\qquad \text{Evolution unitaire non bruitée} \\ &[\sum_i c_i|\phi_i\rangle] \otimes |e'\rangle \\ &\Rightarrow \langle\psi|\mathcal{A}|\psi\rangle = \sum_{i,j} c_i^* c_j \langle\psi_i|A|\psi_j\rangle \underbrace{\langle e'|e'\rangle}_{=1}.\end{aligned} \qquad (2.6)$$

Ainsi, dans le cas d'une évolution bruitée, nous nous retrouvons avec un état de mélange pour le système, alors que dans l'autre cas, nous gardons un état de superposition. Lorsque nous tenons compte du couplage entre le système et l'environnement, il y a une perte d'information. Comme nous l'avons montré précédemment, la perte d'information est complète lorsque les états de l'environnement $|e_i\rangle$ sont orthogonaux. On appelle alors *temps de décohérence* $\tau_{dec}$ le temps typique sur lequel cette perte d'information se produit. Cette décohérence peut également être décrite en ne considérant que le système et en invoquant une réduction du paquet d'onde ("collapse"). Le temps de décohérence s'interprète alors comme le temps qui s'écoule avant que le collapse n'ait lieu. La limitation principale de l'ordinateur quantique réside dans ce temps de décohérence. En effet, suite au collapse, il n'est plus possible de continuer le calcul. Il est donc nécessaire d'avoir réalisé toutes les opérations souhaitées avant que le phénomène n'ait lieu. Nous reviendrons sur cette limitation dans la section 3.2.1.

## 2.2 Réinterprétation en langage informatique

- En considérant des systèmes à deux niveaux, il est possible, au lieu de parler de spin up et down, d'hélicité + et − ou de niveau d'énergie f (fondamental) et e (excité), de parler de *Q-bit* (quantum binary digit) prenant la valeur 0 ou 1 :

$$\left.\begin{aligned}|\uparrow\rangle &, |\downarrow\rangle \\ |+\rangle &, |-\rangle \\ |f\rangle &, |e\rangle\end{aligned}\right\} \longrightarrow |0\rangle, |1\rangle. \qquad (2.7)$$

Par cette simple nomenclature, nous définissons les support de l'information.

- En continuant notre nomenclature, le système physique considéré devient un *ordinateur*. Pour créer un ordinateur travaillant avec $N$ Q-bits, il suffit, par analogie au cas classique, de prendre $N$ systèmes à deux niveaux. Ceci est satisfaisant pour l'esprit, mais n'est pas nécessaire. En effet, un seul système possèdant $2^N$ niveaux convient tout aussi bien du moment qu'un adressage cohérent est effectué. Cependant, pour simplifier la suite de l'exposé, nous traiterons toujours le cas d'un ordinateur quantique composé de $N$ spins $\frac{1}{2}$, ce qui ne restreint pas la généralité des propos.

- Finalement, l'évolution du système se traduit par un calcul à l'aide de *portes logiques*. Les portes logiques élémentaires à l'aide desquelles tout calcul peut être réalisé, sont au nombre de deux : la rotation et l'opération XOR (cf. 2.3.1). Ainsi, le schéma de base est le suivant :

$$
\begin{array}{ccccc}
\text{préparation} & - & \text{évolution} & - & \text{détection} \\
\downarrow & & \downarrow & & \downarrow \\
\text{entrée} & - & \text{calcul} & - & \text{sortie.}
\end{array}
\tag{2.8}
$$

*Remarque :* Dans le langage des machines de Turing, le système joue le rôle de la "bande infinie" sur laquelle l'information est lue et enregistrée, et l'évolution joue naturellement le rôle du processeur.

A ce stade, nous possédons des Q-bits, un ordinateur avec une entrée et une sortie, ainsi que la possibilité d'implémenter n'importe quelle opération unitaire. Il ne manque donc plus que des algorithmes intéressants pour que l'ordinateur quantique soit théoriquement achevé.

## 2.3 L'algorithme

### 2.3.1 Définition du concept de l'algorithme

Un *algorithme* est une suite convenable d'évolutions unitaires et de mesures sur le système. On veut pouvoir effectuer n'importe quelle opération unitaire sur le système de $N$ Q-bits. Il a été montré (Barenco et al. [8]) que toute opération unitaire peut être construite comme le produit de deux opérations simples :

- **La rotation R.** Cette opération sert à changer un seul bit indépendamment des autres. Comme par convention les spins up et down se situent sur les directions $+z$ et $-z$, cette opération unitaire est de la forme :

$$\begin{aligned} R_{\theta\phi}|0\rangle &= \cos\theta|0\rangle + e^{-i\phi}\sin\theta|1\rangle \\ R_{\theta\phi}|1\rangle &= -e^{i\phi}\sin\theta|0\rangle + \cos\theta|1\rangle \end{aligned} \quad (2.9)$$

où $\theta$ et $\phi$ sont les angles azimutal et polaire respectivement.

Mise à part la difficulté de pouvoir appliquer pratiquement cette opération sur chaque Q-bit séparément, il est intéressant de remarquer que celle-ci donne lieu à un état de superposition, concept appartenant uniquement à la physique quantique. Cette opération est donc intrinsèquement non-classique.

- **L'opération XOR.** L'opération XOR (exclusive OR, appelée aussi CNOT : controlled-NOT) est une opération logique entre deux Q-bits : elle agit sur un Q-bit donné selon l'état d'un autre Q-bit. Soit $|\sigma_n\sigma_m\rangle$ l'état de deux Q-bits quelconques repérés par $n$ et $m$. L'opération XOR, notée $C(n,m)$, peut se définir ainsi :

$$\begin{aligned} C(n,m)|00\rangle &= |10\rangle \\ C(n,m)|10\rangle &= |00\rangle \\ C(n,m)|01\rangle &= |01\rangle \\ C(n,m)|11\rangle &= |11\rangle. \end{aligned} \quad (2.10)$$

L'état du Q-bit $n$ est changé si et seulement si l'état du Q-bit $m$ est $|0\rangle$. L'opération XOR couple donc deux Q-bits quelconques mais, contrairement à la rotation, reste une opération classique [2]. Comme nous le voyons dans l'exemple suivant, l'effet de cette opération dans un ordinateur quantique porte sur le degré d'intrication des états :

$$\begin{aligned} C(n,m)\left[|00\rangle + |11\rangle\right] &= |10\rangle + |11\rangle \\ &= |1\rangle \otimes \left[|0\rangle + |1\rangle\right]. \end{aligned} \quad (2.11)$$

A l'aide de la rotation, nous pouvons donc changer n'importe quel Q-bit indépendamment des autres et avec l'opération XOR, nous pouvons agir sur le degré d'intrication des états. Ainsi, tout l'espace de Hilbert est atteint. Pour illustrer la manipulation de ces opérations, prenons un système à 3 spins. Nous pouvons, par exemple, à partir de l'état $|111\rangle$ obtenir un état d'intrication maximale:

$$\begin{aligned} \left[\mathbb{1} \otimes \mathbb{1} \otimes R\left(\tfrac{\pi}{4}\right)\right]|111\rangle &= \tfrac{1}{\sqrt{2}}(|111\rangle + |110\rangle) \\ \left[\mathbb{1} \otimes C(2,3)\right]\tfrac{1}{\sqrt{2}}(|111\rangle + |110\rangle) &= \tfrac{1}{\sqrt{2}}(|111\rangle + |100\rangle) \\ \left[C(1,2) \otimes \mathbb{1}\right]\tfrac{1}{\sqrt{2}}(|111\rangle + |100\rangle) &= \tfrac{1}{\sqrt{2}}(|111\rangle + |000\rangle). \end{aligned} \quad (2.12)$$

Pour d'autres exemples didactiques, il est conseillé de consulter l'article de V. Scarani [9].

---

[2] Toute la logique classique peut être réalisée à partir de portes XOR.

### 2.3.2 Avantage d'un algorithme quantique

A l'heure actuelle, quelques algorithmes quantiques plus performants que leurs homologues classiques sont connus. La particularité de tous ces algorithmes est l'utilisation d'états superposés. Par ce biais, le nombre d'opérations de base nécessaires est nettement plus faible qu'avec un algorithme classique. Ainsi, l'algorithme de Shor [2] permet de factoriser un nombre donnée à l'aide de $N^\alpha$ opérations, où $N$ est la taille du nombre à factoriser. Les algorithmes classiques pour ce genre de problèmes nécessitent un nombre d'opérations qui croît exponentiellement avec $N$. L'algorithme de Grover [3] qui permet de trouver un élément dans une liste est également moins complexe que son homologue classique. Nous nous rendons alors compte que le gain de temps ne se réalise pas sur la rapidité d'exécution d'une opération de base, mais sur le nombre d'opérations. Par conséquent, ce qui peut définir un ordinateur de quantique est la façon d'utiliser le système mais non pas le système lui-même. Ce n'est pas parce qu'un système de spins est utilisé qu'un ordinateur est quantique.

Il existe d'autres algorithmes qui, tout en étant spécifiquement quantiques, ne changent pas la complexité. Par exemple celui de Fahri et al. (détermination de la parité d'une fonction) nécessite $N/2$ opérations [10], alors que son homologue classique en nécessite $N$. Dans la section suivante, nous illustrons le concept d'algorithme au moyen de l'algorithme de Deutsch-Jozsa[4].

### 2.3.3 Un exemple : l'algorithme de Deutsch-Jozsa

Prenons un système $S$ composé de deux sous-systèmes $S_1$ et $S_2$. Il est ainsi possible de construire les quatre fonctions différentes agissant sur l'ensemble $\{0,1\}$ :

$$f_1(x) = 0, \quad f_2(x) = 1, \quad f_3(x) = x, \quad f_4(x) = NOTx \tag{2.13}$$

L'algorithme de Deutsch-Jozsa permet de savoir si une fonction prise au hasard est constante ($f_{1,2}$) ou non ($f_{3,4}$). Pour ce faire, quatre étapes suffisent :

1. Préparation de l'état initial, en l'occurence : $|0\rangle \otimes |0\rangle$

2. Rotation de l'état de chaque spin, livrant ainsi un état de superposition :

$$\tfrac{1}{2}(|0\rangle + |1\rangle) \otimes (|0\rangle - |1\rangle) = \tfrac{1}{2}\sum_{x=0}^{1} |x\rangle \otimes (|0\rangle - |1\rangle) \tag{2.14}$$

3. Appel de $f(x)$. Il faut appliquer $(1 + f(x))\,mod\,2$ sur le système $S_2$. Remarquons que si $f = f_{3,4}$, cette étape nécessite l'utilisation de l'opération XOR. Le système $S$ se trouve alors

dans l'état:

$$\tfrac{1}{2}\sum_{x=0}^{1}|x\rangle\otimes\underbrace{(|(0+f(x))\,mod2\rangle-|(1+f(x))\,mod2\rangle)}_{=(-1)^{f(x)}(|0\rangle-|1\rangle)}=$$

$$=\tfrac{1}{2}\left(\sum_{x=0}^{1}(-1)^{f(x)}|x\rangle\right)\otimes(|0\rangle-|1\rangle)=\begin{cases}\pm\tfrac{1}{2}(|0\rangle+|1\rangle)\otimes(|0\rangle-|1\rangle) & \text{si } f=f_{1,2}\\ \pm\tfrac{1}{2}(|0\rangle-|1\rangle)\otimes(|0\rangle-|1\rangle) & \text{si } f=f_{3,4}\end{cases} \quad (2.15)$$

4. Rotation inverse de celle décrite au point 2 pour trouver l'état final du système :

$$\begin{aligned}|0\rangle\otimes|0\rangle & \text{ si } f=f_{1,2}\\ |1\rangle\otimes|0\rangle & \text{ si } f=f_{3,4}\end{aligned} \quad (2.16)$$

La mesure du premier Q-bit nous indique ainsi si la fonction est constante ($|0\rangle$) ou non ($|1\rangle$).

Le point remarquable de cet algorithme est que la fonction $f$ que nous désirons tester n'est appelée qu'une seule fois, alors que dans un algorithme classique deux appels sont nécessaires. A la section 3.2.1, nous décrirons une réalisation pratique de cet algorithme à l'aide de la résonance magnétique nucléaire.

## 2.4 Conclusion

Au terme de cette section, nous remarquons donc qu'un ordinateur quantique est n'importe quel système quantique vu comme porteur d'information. Au niveau théorique, il n'y a rien de nouveau. Nous assistons seulement à une réinterprétation de concepts déjà établis en mécanique quantique. Pour pouvoir effectuer un calcul quantique, il faut se donner la possibilité d'effectuer n'importe quelle opération unitaire. Pour ce faire, deux opérations suffisent : la rotation et l'opération XOR. Cependant, le couplage du système avec l'environnement pose de graves problèmes : il faut donc réaliser toutes les opérations nécessaires pour effectuer le calcul avant l'échéance de $\tau_{dec}$ ($\Rightarrow$ limitation du nombre d'opérations) ou alors trouver un moyen de vaincre ce phénomène. Ainsi, dans toute proposition de réalisation d'un ordinateur quantique, nous trouverons trois éléments : le système physique envisagé, la manière d'implémenter les deux opérations et la manière de vaincre la décohérence.

Soulignons encore que le point crucial d'un algorithme quantique est l'utilisation de la superposition d'états, c'est-à-dire que l'information elle-même est quantique. Un système quantique qui traite de l'information classique n'est pas un ordinateur quantique. En effet, les semiconducteurs actuels possèdent déjà un caractère quantique bien que les algorithmes utilisés soient classiques. Notons également que, si le système physique est composé de sous-systèmes, alors les états intriqués seront également essentiels, car ils sont une conséquence directe de la superposition. Cependant, si nous considérons uniquement le système global, nous verrons apparaître des états superposés, mais aucun état intriqué. Ainsi, la notion d'état intriqué dépend du point de vue que l'on adopte, mais l'état physique réel auquel il fait référence demeure essentiel si nous désirons utiliser toutes les possibilités d'un ordinateur quantique.

# Chapitre 3

# Réalisations pratiques

## 3.1 Les différents systèmes utilisés

Différents systèmes physiques sont considérés pour la réalisation pratique. Les principaux sont les suivants [5] :

- **Les ions piégés**. Un ion piégé peut être vu comme une particule plongée dans un potentiel (problème hydrogénoïde). Il est alors possible de considérer deux configurations différentes du ion, donc deux niveaux d'énergie distincts. A l'aide d'impulsions laser nous pouvons modifier l'état de chaque ion et l'intrication des états. Par cette technique, le groupe de Boulder a obtenu une intrication de quatre Q-bits [11].

- **Les boîtes quantiques** (quantum dots). Les boîtes quantiques sont en fait une réalisation des puits de potentiel carré souvent considérés dans la théorie. Le Q-bit fait alors référence à deux niveaux d'énergie différents d'un électron placé dans un de ces puits. A ma connaissance, aucun état intriqué n'a pu être réalisé jusqu'à présent.

- **Les photons**. Dans le cas des photons, les deux états du Q-bit sont deux états de polarisation et les opérations sont réalisées par d'astucieuses techniques d'interférométrie. L'intrication à trois photons a été démontrée. Comme le support de l'information n'est pas solide, l'interférométrie de photons est plutôt envisagée pour la communication et la cryptographie quantiques.

La seule technique qui a réussi à implémenter des algorithmes, notamment celui de Deutsch-Jozsa et un code de correction d'erreurs, est la résonance magnétique nucléaire, que nous analysons plus en détail dans la prochaine section.

## 3.2 Réalisation par résonance magnétique nucléaire

### 3.2.1 L'expérience de Chuang et al.

De nombreux groupes de recherche tentent de réaliser un ordinateur quantique à l'aide de la résonance magnétique nucléaire (RMN). La principale question qui se pose est la suivante : comment travailler avec des états purs [1] dans des expériences de RMN effectuées à température ambiante, donc avec des états distribués thermiquement ? La réponse se trouve dans la notion d'états pseudo-purs, introduite en 1997 par Cory et al. [12] et par Gershenfeld et Chuang [13].

Considérons un système $S$ composé de deux spins. En RMN, ces spins sont plongés dans un champ magnétique extérieur. Les états propres de cette interaction sont $|++\rangle, |+-\rangle, |-+\rangle, |--\rangle$. A l'équilibre, la probabilité d'occupation de ces états est donnée par la statistique de Boltzmann :

$$\begin{aligned} P_{\sigma\omega} &= \frac{e^{\frac{E_{\sigma\omega}}{kT}}}{Z} \qquad \sigma, \omega = +, - \\ Z &= \sum_{\sigma,\omega} e^{\frac{E_{\sigma\omega}}{kT}} \quad \text{fonction de partition.} \end{aligned} \qquad (3.1)$$

Nous nous rendons alors compte que si $T \to \infty$ chaque état est peuplé équitablement ($P_{\sigma,\omega} = \frac{1}{4}, \forall \sigma, \omega$). Les écarts typiques à la distribution uniforme sont de l'ordre de $\frac{E}{kT} \approx 10^{-5}$ à température ambiante. Il est habituel en RMN d'écrire la matrice densité totale $\rho_t$ comme :

$$\rho_t = \frac{1}{4}\mathbb{1} + \rho. \qquad (3.2)$$

L'intérêt de cette décomposition vient du fait que $\rho$ donne les seules contributions non-triviales à la dynamique. La matrice $\rho$ est appelée matrice densité réduite et nous remarquons qu'elle doit être de trace nulle puisque $\text{Tr}(\rho_t) = 1$. Cette matrice densité réduite ne représente donc pas un état (la trace de toute matrice densité représentant un état, pur ou de mélange, vaut 1). Cependant, on a montré l'existence d'une préparation $U$ astucieuse (séquences d'impulsions en RMN) telle que:

$$\begin{aligned} U^+ \rho_t U &= \tfrac{1}{4}\mathbb{1} + U^+\rho U \\ &= \tfrac{1}{4}\mathbb{1} - \tfrac{\epsilon}{4}\mathbb{1} + \underbrace{\tfrac{\epsilon}{4}\mathbb{1} + U^+\rho U}_{=\epsilon\rho_1} \\ &= \left(\tfrac{1-\epsilon}{4}\right)\mathbb{1} + \epsilon\rho_1, \end{aligned} \qquad (3.3)$$

avec $\rho_1$ la matrice densité décrivant un état pur. Le terme $\left(\frac{1-\alpha}{4}\right)\mathbb{1}$ ne va pas influencer la dynamique du système. Ainsi, l'analyse de $\rho_1$ suffit. On parle alors d'états pseudo-purs. De cette manière, les chercheurs justifient leur tentative de réaliser un ordinateur quantique à l'aide de la RMN. Ce résultat se généralise facilement pour un système de $N$ spins :

$$\rho_\epsilon = \tfrac{1-\epsilon}{d}\mathbb{1}_d + \epsilon\rho_1, \quad d = 2^N \tag{3.4}$$

---

[1] Il ne suffit pas que le système ait une évolution bien déterminée, faut-il encore que nous puissions lire le résultat et l'interpréter.

Cette justification théorique étant explicitée, nous pouvons nous intéresser à la réalisation pratique de l'algorithme de Deutsch-Jozsa (2.3.3) effectué par I.L. Chuang et al. [14]. Les deux Q-bits sont les spins nucléaires du carbone et de l'hydrogène dans une molécule de chloroforme ($CHCl_3$). Comme ces spins ont des fréquences de résonance différentes, il est aisé d'effectuer une rotation sur l'un des systèmes, sans influencer le second. L'opération XOR se réalise à partir d'une judicieuse combinaison d'impulsions (techniques de double résonance). Nous n'entrerons pas ici dans les détails techniques des séquences utilisées. En effectuant les quatre étapes proposées par l'algorithme, le pic de résonance apparaît selon $+z$ pour les fonctions $f_{1,2}$ et selon $-z$ pour les fonctions $f_{3,4}$, comme attendu par la théorie.

Nous pourrions alors louer la réussite. Remarquons cependant que ce genre de techniques RMN sont à l'heure actuelle maîtrisées [2] et que l'algorithme était déjà établi. De plus, comme le nombre d'opérations est assez restreint pour le cas traité, les auteurs n'ont pas eu à se préoccuper du problème majeur : la décohérence. En effet, dans ce cas très simple, le couplage entre le système et l'environnement n'a pas le temps d'agir [3]. Cependant, dès que le problème se complique légèrement, les choses deviennent catastrophiques.

Pour illustrer ces propos, nous suivons l'article de Haroche et Raimond [15]. Rappelons que dans la section 2.1.3 nous avons défini la grandeur $\tau_{dec}$ comme étant le temps disponible pour effectuer nos opérations, avant que n'ait lieu la réduction du paquet d'onde. Ainsi, connaissant le temps nécessaire pour effectuer une opération ($\tau_{op}$), nous pouvons définir la grandeur $M$, le nombre de pas qu'il est possible de réaliser sans perdre d'information.

$$\begin{aligned}M &= \tfrac{\tau_{dec}}{\tau_{op}} \\ M &\approx 10^7\end{aligned} \tag{3.5}$$

---

[2] Les premières expériences de double irradiation ont été effectuées par Bloch en 1954 ; dès la fin des années 50, ces techniques apparaissent dans les livres

[3] Dans ces expériences de RMN, il est habituel de considérer le temps de relaxation $T_2$ comme le temps de décohérence; pour l'heure, il s'agit d'une estimation d'ordre de grandeur plutôt que d'une idenfication conceptuelle entre $T_2$ et $\tau_{dec}$ (e-mail de R. Laflamme à V. Scarani)

Pour espérer factoriser un nombre de 4 bits, $10^6$ opérations sont nécessaires et pour un nombre de 400 bits, $10^{12}$ opérations. En se rappelant que le plus grand nombre de 4 bits est 15, l'ordinateur quantique semble assez limité. Pour remédier à ce problème, des protocoles de corrections sont mis au point. L'idée théorique de ces codes est assez simple. Il suffirait de déterminer l'opérateur unitaire d'évolution bruitée, d'en calculer l'inverse, et de l'appliquer sur l'état. Ainsi, la donnée serait maintenue dans son état initial. Nous ne sommes cependant pas au bout de nos peines, car un algorithme de correction est complexe et nécessite également la superposition d'état. Un deuxième ordinateur quantique serait donc nécessaire...

Ainsi, malgré sa publication dans la prestigieuse revue *Nature*, l'expérience de Chuang et al. n'est pas aussi révolutionnaire qu'il n'y paraît au premier abord : tous les éléments théoriques sont connus, la technique est maîtrisée et les problèmes pratiques intéressants à résoudre sont contournés. Pour couronner le tout, un article de S.L. Braunstein et al. [16] démontre la proposition suivante: *"Tous les états utilisés jusqu'à maintenant en RMN pour l'ordinateur quantique ou pour d'autres protocoles d'information quantique sont séparables" (i.e. non intriqués).* L'idée de la démonstration est la suivante. Le premier pas est une décomposition de la matrice densité définie par l'équation (3.4) dans une base surdimentionnée, en l'occurence à l'aide des matrices de Pauli. Le critère suivant est alors appliqué: *si tous les coefficients de la décomposition sont non négatifs, alors la matrice considérée est séparable.* Comme les coefficients négatifs proviennent uniquement de la matrice $\rho_1$, si le coefficient $\epsilon$ est suffisamment petit, alors la matrice densité est séparable. En calculant une borne inférieure pour ce coefficient, ces auteurs démontrent que dans toute les tentatives de réalisation d'ordinateur quantique par RMN, les états intriqués physiques ne sont pas accessibles. L'état pseudo-pur est un simulateur d'état pur. Ainsi, bien que l'expérience réalisée par Chuang et al. utilise un algorithme quantique, ce dernier ne nécessite pas forcément l'utilisation d'états intriqués contrairement à l'algorithme de Shor. Un ordinateur réalisé par RMN à des températures "normales" ne pourra donc jamais factoriser un nombre donné avec le nombre d'opérations minimales prévues. Quelle est donc la véritable définition d'un ordinateur quantique ? A ce stade de la recherche naîssent des confusions ainsi qu'un débat de fond.

### 3.2.2 Le débat de l'ordinateur quantique par résonance magnétique nucléaire

Ce débat est plus qu'une simple discussion entre connaissseurs du domaine. Dans ce sens, l'article de *Physics Today* [17] nous dépeint le tableau actuel à l'aide d'interviews des différents opposants. Suite à sa démonstration de la séparabilité de la matrice densité des états pseudo-purs en RMN,

S.L. Braunstein pose la question de base suivante : *"Lorsque nous pensons à l'ordinateur quantique, nous pensons que c'est merveilleux, et qu'il utilise les fantastiques propriétés d'un système quantique. Ces propriétés sont des choses comme la superposition et l'intrication. Ainsi, quel sens ont ces machines si elles ne peuvent produire aucun état intriqué ?"* Selon S. Popescu, l'intrication est l'élément de base pour le succès de l'ordinateur quantique. Ainsi, cette démonstration implique de sérieux doutes sur les réalisations d'ordinateur quantique par RMN. Les expérimentateurs qui travaillent sur ce projet se défendent d'une part par l'unitarité des transformations et d'autre part par l'évolution du système. Selon R. Laflamme, le fait d'utiliser des tranformations unitaires (à caractère quantique) permet d'obtenir la réponse cherchée de manière plus efficace. De plus, Il n'a pas été possible à ce jour de décrire l'évolution observée de manière purement classique. Cependant, Linden et Popescu ont montré que l'algorithme de factorisation de Shor nécessite l'utilisation d'états intriqués, et par conséquant, une évolution quantique ne suffit pas. Les ordinateurs réalisés par RMN ne pourront donc jamais atteindre la puissance espérée d'un ordinateur quantique. Il semble donc que ces machines ne soient qu'une "simulation" d'un véritable ordinateur quantique. C'est dans cette brêche que s'engouffre S. Lloyd en affirmant qu'il existe une multitude de façon d'implémenter des opérations, et pas seulement de façons classique ou quantique. Le fait que l'ordinateur construit par RMN se trouve à mi-chemin entre les mondes classique et quantique devient ainsi simultanément un chef d'accusation et une défense. Devant de telles discussions, il est légitime de se demander si le débat porte sur des faits scientifiques ou si nous assistons simplement à un problème de linguistique. Il ne serait pas impossible que dans un proche avenir une définition plus stricte de l'ordinateur quantique apparaisse.

Lorsque nous quittons la presse spécialisée pour des revues destinées à un plus large publique, nous nous rendons compte que le flou linguistique est encore plus prononcé. Prenons comme exemple l'article de J. Dousson paru dans le *Flash Informatique* de l'EPFL [18]. Dans son introduction, elle nous explique, graphique à l'appui, que d'ici 2020, quelques atomes suffiront pour stocker un bit. Peut-être. Alors continue-t-elle, sur de tels systèmes, seules les lois de la mécanique quantique sont valables. Sûrement. Par consèquent, nous travaillerons avec de l'informatique quantique. Sûrement pas. Nous retrouvons ici de façon flagrante la confusion entre un système quantique et de l'information quantique. En dehors de ces confusions, il est quand même intéressant de souligner que ces articles de vulgarisation se multiplient dans les différentes revues (par ex. *l'Ordinateur Individuel*) et même dans les quotidiens (par ex. *La Liberté*).

Au terme de ce chapitre, nous nous rendons compte que la réalisation pratique est non seulement difficile, mais qu'une fois un prototype réalisé, il n'est pas évident de savoir (ou d'admettre) s'il remplit toutes les exigences requises pour que nous puissions parler d'ordinateur quantique.

# Chapitre 4

# Aspect social de l'ordinateur quantique

## 4.1 Les publications

### 4.1.1 Explosion de la recherche

Comme nous l'avons déjà vu en introduction, c'est en 1994 que P. Shor trouva le premier algorithme intéressant pour l'ordinateur quantique. Par cette découverte, l'idée qu'un ordinateur quantique permet de résoudre des problèmes plus rapidement qu'un ordinateur classique était née et par la même occasion, une explosion de la recherche dans ce domaine.

A l'aide de la base de données INSPEC (Intranet de l'EPFL : biberl.epfl.ch/cgi-bin/webspirs.cmd), le nombre d'articles publiés par année et qui contiennent l'un des syntagmes "quantum computer", "quantum computing", "quantum computation" a été évalué. Comme indicateur de la qualité des articles, une seconde courbe ne comprenant que les articles parus dans *Physical Review* a été ajoutée. Les résultats sont présentés sur la figure 4.1. Il est à signaler que, lors d'un survol, environ 5% de ces articles ne traitent pas de l'ordinateur quantique, le syntagme "quantum computation" pouvant être utilisé pour décrire un calcul numérique d'un système quantique. Sur la figure réalisée, il est aisé de constater une explosion de la recherche dans ce domaine dès 1994, donc après l'article de P. Shor. Nous nous rendons également compte que le quart des articles ont été publiés dans *Physical Review*.

Par la lecture des abstracts, nous constatons que les années 1997-98 correspondent à l'achèvement de la théorie et à quelques idées de réalisations pratiques. En effet, les scientifques reconsidèrent les algorithmes et s'intéressent aux corrections nécessaires pour minimiser la perte d'information par la décohérence (cf. 2.1.3). En 1999, les notions théoriques étant quasiment complètes, ce sont des tentatives pratiques qui dominent la recherche.

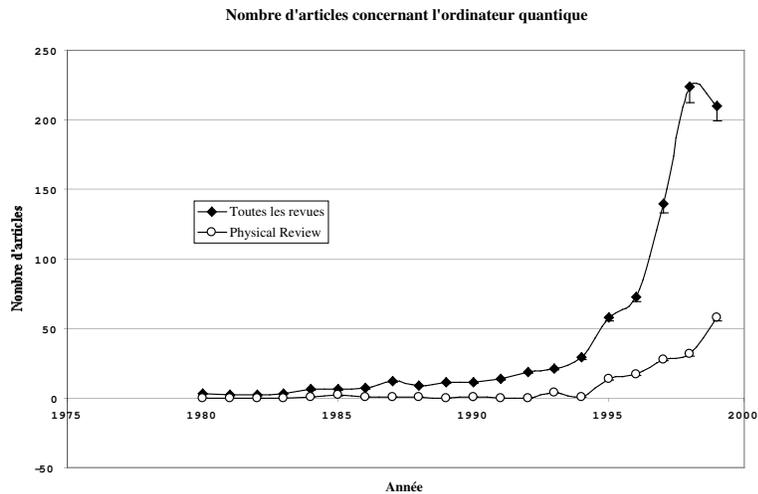

Figure 4.1: Explosion de la recherche sur l'ordinateur quantique

Quelles peuvent être alors les raisons de cet engoûment pour l'ordinateur quantique ? Avec 210 articles parus en 1999, donc plus de 4 par semaine, alors que le sujet n'était quasiment pas traité 5 ans auparavant, nous pouvons nous demander si l'ordinateur quantique est aussi prometteur que la figure 4.1 nous le suggère, ou si nous sommes en présence d'un effet de mode.

### 4.1.2 Raisons

Nous proposons ici une analyse des différentes raisons qui peuvent conduire un groupe de scientifiques à se lancer dans cette recherche.

La première est tout simplement la *passion du sujet*. Dès son plus jeune âge, la mécanique quantique a suscité de grands débats. La discrétisation des niveaux d'énergie, la dualité onde-corpuscule ont provoqué un doute même chez les plus grands de l'époque comme Einstein. Dans cette continuité, un calcul sur un ordinateur quantique débute avec un état connu pour terminer avec un état bien défini, en passant par un état de superposition. Le point intrigant de ce schéma réside dans le fait qu'à un certain moment du calcul, nous ne pouvons pas connaître l'état dans lequel se trouve le système. En effet, considérons uniquement deux spins et supposons que le système global se trouve dans l'état $\frac{1}{\sqrt{2}}[|11\rangle + |10\rangle]$. Nous avons une probabilité $\frac{1}{2}$ que, lors d'une mesure, le système soit dans l'état $|11\rangle$ et également $\frac{1}{2}$ qu'il soit dans l'état $|10\rangle$. De plus, si nous effectuons la mesure, nous avons une modification de l'état, donc une perte d'information. L'ordinateur quantique entre donc dans la lignée de ces éléments surprenants, mais toujours passionnants, de la mécanique quantique.

La seconde est l'*utilité* de l'application finale. L'ordinateur quantique, mais aussi la cryptographie ou la théorie de l'information quantiques, semblent être les premières démonstrations de l'utilité de la superposition d'états, notion qui relevait jusqu'à très récemment des débats conceptuels. En effet, comme cette dernière permet d'établir des algorithmes plus rapides, nous pouvons espérer découvrir la solution de problèmes jusqu'alors qualifiés d'irrésolvables ou d'approximatifs de par la limitation de nos outils informatiques. Par exemple, la cryptographie classique, qui se base sur la factorisation d'un nombre donné en deux nombres premiers, deviendrait désuète.

Une troisième raison pourrait s'appeler le *recyclage scientifique*. Depuis 1960, de nombreuses personnes se sont intéressées aux fondements de la mécanique quantique, notamment à certains concepts comme la non-localité ou l'existence de variables cachées [19]. Cependant, à l'heure actuelle, les expériences réalisables sont de plus en plus nombreuses et en bon accord avec les prédictions de la théorie. Une réinterprétation de la mécanique quantique permet alors d'éviter de mettre un point final prématurément à son élaboration.

Finalement, comme presque toujours, l'*argent* est une excellente raison. Une entreprise qui commercialiserait un ordinateur surpassant tous les modèles existants réaliseraient certainement de gigantesques bénéfices. Ainsi, en lui faisant miroiter une telle application, les scientifiques du domaine s'assurent les fonds nécessaires pour leur recherche. Soulignons également le fait qu'une partie importante du financement provient d'organismes nationaux ou supranationaux. En effet, de nos jours, la possibilité d'application est un critère de financement non seulement pour les privés, mais aussi pour ces organismes. Par exemple, le projet TOP NANO 21 du Fonds National Suisse exige un partenariat industriel. L'ordinateur quantique peut ainsi satisfaire un critère de rentabilité demandé par les privés mais malheureusement aussi par les organismes nationaux.

Toutes les raisons mentionnées — la passion, l'utilité, le recyclage scientifique et l'argent — me paraissent être de bonnes motivations pour se lancer dans cette recherche. Cependant, comme nous l'avons vu dans la section précédente, la grande limitation pratique et le peu d'algorithmes existants me poussent à croire que ce sont les deux dernières les principales.

## 4.2 L'ordinateur quantique en Suisse et dans le monde

Peu de groupes de recherche en Suisse s'occupent de l'ordinateur quantique. Le seul qui travaille activement dans le domaine (ordinateur quantique à l'aide de "quantum dots") est D. Loss, à Bâle (theorie5.physik.unibas.ch). Le groupe de N. Gisin, à Genève, est plus spécialisé en cryptographie et théorie de l'information quantiques qu'en ordinateur quantique (www.gap-optique.unige.ch). Une brève recherche sur Internet montre que les autres sites où figure l'ordinateur quantique ne traitent pas du sujet, mais le placent comme application directe de leur recherche. Dans ce contexte, nous retrouvons notamment les personnes s'occupant de la physique mésoscopiques, par exemple J. Faist, à Neuchâtel (www.unine.ch/uer/uer_physique.htm). L'EPFL n'y échappe pas non plus: l'IMO qui travaille sur les "quantum dots" a récemment invoqué l'ordinateur quantique comme application possible (Polyrama 112, décembre 1999, p. 36-38).

Pour les gens qui s'intéressent à l'ordinateur quantique dans le monde, citons l'adresse internet du Center for Quantum Computation, situé à Oxford : www.qubit.org. Ce site propose la liste des principaux groupes de recherche dans le monde et en Europe. Notons qu'il est sponsorisé par différentes entreprises et organismes, dont, comme par hasard, Hewlett Packard.

# Chapitre 5

# Conclusion

D'après le nombre d'articles publiés par année, l'ordinateur quantique semble être une application très utile et la recherche avancée. Cependant, la réalisation pratique est difficile de par la limitation du nombre d'opérations par la décohérence. Comme nous l'avons vu, $10^7$ opérations sont réalisables avant que la décohérence n'agissent, et $10^6$ (resp. $10^{12}$) sont nécessaires pour factoriser un nombre de 4 (resp. 400) bits. La réalisation d'états intriqués avec un grand nombre de Q-bits est également un grand défi. La factorisation d'un nombre donné $n$ par l'algorithme de Shor exige l'entrée du nombre $\frac{n}{2}$. Ainsi, la factorisation de 30 nécessite 4 Q-bits, celle de 100, 9 Q-bits, celle de $10^6$, 20 Q-bits et pour les applicatons visées, c'est la factorisation de nombres proches de $10^{100}$, donc environ 330 Q-bits, qui est intéressante. Cependant, à l'heure actuelle, le meilleur résultat est la réalisation d'un état intriqué de 4 Q-bits. De plus, nous nous rendons intuitivement compte que plus le nombre de Q-bits est élevé, plus le temps de décohérence est court. Ces ordres de grandeur me semblent alors accablantes pour un réalisation pratique de l'ordinateur quantique. Les optimistes me répondront que, si quelqu'un avait affirmé à Jules César qu'un jour l'homme marcherait sur la lune, il se serait fait dévorer par les lions. D'accord, mais entre le char de Ben-Hur et la Formule 1 d'aujourd'hui, l'humanité a assisté à quelques révolutions technologiques. A mon avis, une révolution supplémentaire est à l'heure actuelle nécessaire pour une réalisation de l'ordinateur quantique.

De plus, en écartant l'exploit que représente la réalisation d'un ordinateur quantique, les possibilités prévues d'une telle machine ne sont pas si révolutionnaires que l'on veut nous faire croire. En effet, en dépit de la recherche intensive des cinq dernières années, on a trouvé seulement trois algorithmes (détermination de la période d'une fonction, recherche d'un élément d'une liste, caractérisation d'une fonction) qui sont théoriquement plus performants que les algorithmes classiques. Les com-paraisons entre un ordinateur classique et un ordinateur quantique ne sont donc pas équitables. Lorsqu'on nous parle de ce dernier, on nous vante à chaque fois le gain d'opérations que nous pouvons obtenir avec l'algorithme de Shor, mais on oublie volontairement de nous rappeler toutes les autres possibilités qu'offre un ordinateur classique. Quelques scientifiques, notamment E. Farhi et al. [10], ont cependant démontré que dans bien des cas, l'ordinateur quantique ne surpasserait pas son homologue classique. Ainsi, avant de s'investir pleinement dans une tentative de réalisation, une recherche plus approfondie d'algorithmes me paraît nécessaire.

Il me semble donc que les tentatives de réalisations d'un ordinateur quantique sont prématurées et sa puissance théorique exagérée. Les personnes concernées, par leurs arguments et leurs premiers résultats, adoptent le langage de la demi-vérité. Cependant, comme dans notre société le progrès et le profit ne sont plus de simples mots mais des buts en soi, les propos de ces chercheurs ravissent le public et les industries. Ces dernières fournissent alors les fonds nécessaires aux premiers, ravis à leur tour. Dans cet engrenage vicieux, il est naturel que monsieur-tout-le-monde attende désespérément la promotion des ordinateurs quantiques chez Interdiscount.

Je pense donc que l'ordinateur quantique est considéré, à l'heure actuelle, comme une "simple" application de la mécanique quantique, alors qu'il est un sujet de recherche fondamentale. Les quelques éléments prometteurs ont suggéré certaines capacités au concept de l'ordinateur quantique, mais ont surtout fait miroiter un marché juteux. Ainsi, de nombreux chercheurs et industriels se sont engouffrés tête baissée dans cette brêche. Un véritable effet de mode s'en est suivi. A l'heure actuelle, l'ordinateur quantique devient en grande partie un prétexte de publication, un moyen d'obtention de fonds, voire une justification de recherche, ceci en oubliant le premier but du concept. En consultant la littérature, nous observons également un éclatement du sujet. Je ne serais donc pas étonné si, d'ici quelques années, le nombre d'articles publiés annuellement devrait diminuer au profit d'études apparentées comme par exemple la cryptographie quantique.

De par le peu d'algorithmes existants, l'ordinateur quantique ne me semble pas être aussi puissant que le monde se l'imagine et un manque de technologie adéquate me fait penser qu'il est actuellement irréalisable. Ainsi, les optimistes qui prévoient un raz-de-marée d'ordinateurs quantiques en 2020 me paraissent plutôt être des rêveurs ou des "politiciens" que des scientifiques. En attendant le progrès technique suffisant pour une réalisation, espérons que l'ordinateur quantique, replacé dans son contexte de recherche fondamentale, nous élargira l'esprit vers de nouveaux horizons intéressants.

# Bibliographie